# Quantum oscillation studies of the nodal line semimetal Ni$_3$In$_2$S$_{2-x}$Se$_x$

M. M. Sharma[1], Santosh Karki Chhetri[1], Gokul Acharya[1], David Graf[2], Dinesh Upreti[1], Sagar Dahal[1], Md Rafique Un Nabi[1,3], Sumaya Rahman[4], Josh Sakon[5], Hugh O. H. Churchill[1,3,4], Jin Hu[1,3,4*]

[1]Department of Physics, University of Arkansas, Fayetteville, AR 72701, USA

[2]National High Magnetic Field Lab, Tallahassee, FL 32310, USA

[3]MonArk NSF Quantum Foundry, University of Arkansas, Fayetteville, Arkansas 72701, USA

[4]Material Science and Engineering Program, Institute for Nanoscience and Engineering, University of Arkansas, Fayetteville, AR 72701, USA

[5]Department of Chemistry & Biochemistry, University of Arkansas, Fayetteville, AR 72701, USA

**Abstract:**

Ternary shandite compounds with the general formula $T_3M_2X_2$ ($T$ = Ni, Co, Rh or Pd; $M$ = Sn, In, or Pb; and $X$ = S or Se) have emerged as a large pool of topological semimetals. This family of compounds hosts different topological phases for various combinations of $T$, $M$ and $X$. This paper reports the observation of quantum oscillations under the high magnetic fields in Ni$_3$In$_2$S$_{2-x}$Se$_x$ single crystals. Angular dependence of oscillation frequency suggests an evolution of the Fermi surface from three-dimensional to two-dimensional on Se substitution for S in Ni$_3$In$_2$S$_2$. The effective mass obtained for each composition by fitting the oscillation amplitude with the Lifshitz-Kosevich formula, shows no significant change, suggesting that the topological phase might be relatively robust against enhanced SOC upon Se doping in Ni$_3$In$_2$S$_2$.

*jinhu@uark.edu

**Introduction:**

Materials with Kagome lattice have been extensively studied in recent years. The arrangement of atoms in Kagome materials leads to numerous interesting properties such as flat bands originating from the destructive interference of the Bloch wave functions, novel topological quantum states, and spin liquid ground states due to geometric frustrations [1–5]. The shandite compounds with a general formula $T_3M_2X_2$ ($T$ = Ni, Co, Rh or Pd; $M$ = Sn, In, or Pb; and $X$ = S or Se) have attracted significant interest after the discovery of the simultaneous existence of ferromagnetism and Weyl cones in $Co_3Sn_2S_2$ [6–13]. In $T_3M_2X_2$ type compounds, the $T$ atom forms the two-dimensional (2D) Kagome lattice. $Co_3Sn_2S_2$ is the most studied material of this family exhibiting many captivating properties such as robust giant anomalous Hall effect [9,14,15], anomalous Nerst effect [16,17] and large non-saturating magnetoresistance (MR) [12,15,18]. The other Co-based shandite i.e., $Co_3In_2S_2$ is a non-magnetic compound with a linear non-saturating MR, which undergoes an antiferromagnetic transition upon slight doping of Fe atoms [19]. The existence of flat bands and topological states in the $T_3M_2X_2$ family have been studied theoretically and experimentally [7,11,13,20–22]. $Pd_3Pb_2X_2$ is predicted to possess type-I Dirac states which can be tuned to type-II Dirac states by applying external strain [11,20]. Additionally, theoretical and experimental studies on Ni-based shandite $Ni_3In_2S_2$ have shown the presence of endless Dirac nodal lines [7]. Those results highlight the $T_3M_2X_2$ family as a fertile ground to explore various tunable topological states.

$Ni_3In_2S_2$ has emerged as a multiband Kagome system, with non-saturating high MR due to high mobility originating from the linear dispersion of electronic bands [7,23–25]. In $Ni_3In_2S_2$, spin-orbit coupling (SOC) has been predicted to play a major role in determining the topological phases, similar to the scenario of weak topological insulators with an even number of Dirac cones in the presence of SOC [7]. $Ni_3In_2S_2$ is non-magnetic, while its sister compound

Ni$_3$In$_2$Se$_2$ shows antiferromagnetic ordering with Ni moments proposed to aligned within the *ab*-plane [8]. Also, it has been reported that the enhanced SOC in Ni$_3$In$_2$Se$_2$ might lead to the destruction of the topological states, as revealed by the evolution of Berry phase and the calculated gapped Dirac crossing [8]. On the other hand, a recent theoretical study on Ni$_3$In$_2$Se$_2$ show endless Dirac nodal lines similar to its sulphur counterpart [26]. Compared to Ni$_3$In$_2$S$_2$, the topological states in Se-based Ni$_3$In$_2$Se$_2$ and the impact of SOC are less explored. In addition to Angle-resolved photoemission spectroscopy as a direct probe [27,28], quantum oscillation also provides an efficient tool to characterize topological state [29–31]. In this work, we performed quantum oscillation studies on the Ni$_3$In$_2$S$_{2-x}$Se$_x$ system under the high magnetic field using various magnetotransport, tunnel diode oscillator (TDO) and torque magnetometry measurement techniques. Our work extends the quantum oscillation to the high field, revealing the evolution of the Fermi surface with Se substitution and the emergence of a two-dimensional (2D) electronic structure, paving a path for future investigation on electronic and topological properties of Ni$_3$In$_2$S$_{2-x}$Se$_x$ system.

**Experiment**

Single crystalline samples of Ni$_3$In$_2$S$_{2-x}$Se$_x$ ($x$ = 0, 0.4, 1.2, 2) were synthesized using a solid-state reaction method. Stoichiometric ratios of Ni, S and Se powders and In pieces were mixed and vacuum sealed in the quartz ampoule. A Muffle furnace is used to heat the ampoule to 1000ºC in 16 hours. The ampoule was dwelled at 1000ºC for 48 hours and then slowly cooled down to 500ºC at a rate of 2º C/h. After this the furnace was switched off and the ampoule was cooled slowly to the room temperature. Single crystals with silvery metal lustre and easily cleavable were obtained.

X-ray diffraction (XRD) pattern of the powdered Ni$_3$In$_2$S$_{2-x}$Se$_x$ samples were collected by a Rigaku x-ray diffractometer. Rietveld refinement of the powder XRD pattern was performed using the Full Prof software to obtain crystallographic information. Compositions of the synthesized single crystals were determined using Energy Dispersive x-ray spectroscopy (EDS). Low field magnetotransport up to 9 T was measured with a Quantum Design Physical Property Measurement System. High field magnetotransport, TDO and torque magnetometry measurements (see Supplementary Materials) were performed by using the 31 T resistive magnet in the National High Magnetic Field Lab (NHMFL), Tallahassee.

**Results and Discussions**

Figure 1(a) shows the Rietveld refined powder XRD pattern for the synthesized Ni$_3$In$_2$S$_{2-x}$Se$_x$ samples. The XRD pattern for all synthesized samples is found to be well-fitted by a shandite structure having a rhombohedral unit cell with $R\bar{3}m$ space group. The substitution of Se atoms on the S site in Ni$_3$In$_2$S$_2$ results in an increased $c$-axis parameter, which is reasonable for the bigger atomic of Se. The lattice parameters obtained from the Rietveld refinement are summarized in Table 1. Figure 1(b) shows the rhombohedral crystal structure of Ni$_3$In$_2$S$_{2-x}$Se$_x$, and the 2D Kagome lattice formed by Ni-In layer is depicted in Fig. 1(c). The stoichiometry of the synthesized samples has been checked using EDS, shown in table 1 along with the Rietveld refined unit cell parameters. All of the synthesized samples are found to have stoichiometry near the desired ratio.

The end compounds of Ni$_3$In$_2$S$_{2-x}$Se$_x$ i.e., Ni$_3$In$_2$S$_2$ and Ni$_3$In$_2$Se$_2$, have been reported to show strong de Haas-van Alphen (dHvA) oscillations in magnetization measurements starting from a magnetic field as low as 5 T [8,23,26]. Quantum oscillations are a powerful tool to study electronic properties and provide an efficient approach to characterize the topological states [29]. In this work, we extend the dHvA oscillations study to the substituted compositions under a high magnetic field using torque magnetometry and TDO measurements. The Magnetic torque measurements employ a piezoresistive cantilever and measure the torque exerted on a magnetized sample [32]. The magnetic torque depends upon the direction of the applied magnetic field that is easily tunable in NHMFL setup, this makes torque measurement a powerful tool to measure the small anisotropies in the measured system. The oscillating magnetization leads to an oscillating torque signal, which has been widely used in probing electronic structures in topological quantum materials [33–36]. In TDO measurements, a tunnel diode oscillates at a very high frequency, allowing it to detect small changes in the magnetic susceptibility and conductivity [37]. This capability is especially useful for detecting quantum oscillations in materials with very low magnetic moments and conductivity [38,39]. Additionally, as a cross-check, we also adopt Shubnikov-de Haas (SdH) oscillations in magnetotransport enabled by high magnetic fields, which offers a comparison with dHvA effects.

Figure 2 presents high field dHvA oscillations in magnetic torque (Figs. 2a-b) and TDO (Figs. 2c-d) measurements for $Ni_3In_2S_2$ and $Ni_3In_2Se_2$, which was extracted by subtracting a non-oscillating background. The measurements were carried out at different field orientations (see measurement schematic in the inset of Fig. 2a). For magnetic torque oscillation, the oscillation patterns for both samples are complicated, which indicates the presence of multiple frequency components as revealed by fast Fourier transform (FFT) shown in Figs. 2(e) and 2(f) for $Ni_3In_2S_2$ and $Ni_3In_2Se_2$ respectively. For $Ni_3In_2S_2$, for perpendicular field orientation ($\theta = 0°$), FFT analysis shows the presence of five major oscillation frequencies at 57 T, 631 T, 745 T, 990 T and 1213 T, which are indicated as $F_1$, $F_2$, $F_3$, $F_4$ and $F_5$ respectively in Fig. 2e. These frequencies closely match with the frequencies observed in the previous low-field quantum oscillation studies on $Ni_3In_2S_2$ [7,23]. The information of Fermi surface morphology can be obtained by varying field orientations. Evolution of oscillation pattern (Fig. 2a) and frequencies (Fig. 2e) with varying field angles from -5° to 90° is observed, indicating a complicated 3D Fermi surface for $Ni_3In_2S_2$ that is consistent with the previous low field quantum oscillation study [23]. For $Ni_3In_2Se_2$, similarly, three frequencies can be clearly resolved by FFT, $F_1$ = 114 T, $F_2$ = 573 T and $F_3$ = 1346 T [Figs. 2(f)]. The frequencies $F_1$ and $F_3$ are consistent with those extracted from the previous low-field magnetization measurements [8], while our high-field study resolves a clear new frequency $F_2$. In this study, we go beyond the previous study on $Ni_3In_2Se_2$ that focuses on the perpendicular field direction (i.e., $\theta$ = 0°) [8] to examine the angular dependence of quantum oscillation. Similar to $Ni_3In_2S_2$, the first two frequencies $F_1$ and $F_2$ do not vary with the rotation of the magnetic field. Nevertheless, $F_3$ exhibits a $1/\cos\theta$ dependence and is hardly observable when $\theta$ > 30°. Because the quantum oscillation frequency is proportional to the extremal Fermi surface cross-section area perpendicular to the magnetic field direction, such an angular dependence suggests a 2D character for the $F_3$ band. Additionally, as shown in Figs. 2(f), the FFT amplitude (FFTA) of $F_1$ oscillation component of $Ni_3In_2Se_2$ displays non-monotonic angular dependence, which is strong for $\theta$ near 0° and 90° but nearly invisible around $\theta$ ~ 40 - 50°. Such behaviour is in sharp contrast with $Ni_3In_2S_2$ [Figs. 2(e)], which might be associated with the different Fermi surface morphology of the two compounds. Furthermore, it might also be related to the nature of magnetic torque signal, which depends strongly on the angle between the induced magnetic moment and the applied magnetic field.

TDO measurements, which have not been reported for this material family, provide a cross-check to the magnetic torque measurements. The TDO oscillation patterns for $Ni_3In_2S_2$ and $Ni_3In_2Se_2$ have been obtained by removing the non-oscillatory background and are shown in Fig. 2(c) and 2(d) respectively. Similar to magnetic torque measurements, oscillations in TDO also show the presence of multiple frequency components in both $Ni_3In_2S_2$ and $Ni_3In_2Se_2$. FFT analysis reveals multiple frequencies of 114 T, 290 T, 660 T, 804 T, 947 T, 1213 T, and 1981 T for $Ni_3In_2S_2$ under a perpendicular field configuration ($\theta = 0°$), as shown in Fig. 2(f). These frequencies match with the ones observed in torque measurements except for the two additional frequencies at 290 T, and 1981 T. This might be attributed to the fact that the mechanical bending of the cantilever is needed to generate the torque signal. Consequently, the mechanical properties of the cantilever may limit the detection sensitivity of magnetic torque for probing quantum oscillations. Hence, the frequencies 114 T, 660 T, 804 T, 947 T, and 1213 T have been identified as $F_1$, $F_2$, $F_3$, $F_4$ and $F_5$ respectively as observed in torque measurements. For $Ni_3In_2Se_2$, FFT analysis shows the presence of three frequencies $F_1$, $F_2$ and $F_3$ at 114 T, 573 T and 1319 T. The oscillation frequency observed at 1147 T is identified as the second harmonic of $F_2$. The observed oscillation frequencies agree with those probed in torque measurements. In addition, as observed in torque measurements, the higher frequency $F_3$ for $Ni_3In_2Se_2$ shifts towards higher values as rotating magnetic field away from the normal direction, as indicated by the dashed lines in Fig. 2h.

The recent discovery of SdH oscillations in $Ni_3In_2S_2$ has inspired us to perform high-field magnetotransport measurements on $Ni_3In_2S_2$ and $Ni_3In_2Se_2$ single crystals. The SdH oscillations in $Ni_3In_2Se_2$ have not been reported yet as per our knowledge. Figure 3(a) and 3(d) show the magnetoresistance for $Ni_3In_2S_2$ and $Ni_3In_2Se_2$ respectively. The data has been taken at 1.5 K with various magnetic field orientations, as shown in the inset of Fig. 3(a). For $Ni_3In_2S_2$, the oscillations are more pronounced at angles around 30º, and disappear at higher angles above 60º. Similar behaviour is also observed for $Ni_3In_2Se_2$, which is distinct from the clear dHvA effects for the entire magnetic field orientation from 0º to 90º. The observed frequencies are in good agreement with the reported SdH oscillation frequencies [23]. The suppressed SdH effect when magnetic field is rotating approaching to the current direction has also been observed in other materials [40–42] which might be related to the distinct mechanism of SdH and dHvA oscillation: while dHvA oscillation directly probes the oscillating free energy, SdH effect is more complicated as it is sensitive to both density of states and scattering rate [29,43,44]. Nevertheless, for the magnetic field orientations that SdH effect is strong, the resolved oscillation frequencies and their field-angular dependence (Fig. 3c and Fig. 3f) are consistent with these in magnetic torque and TDO oscillations discussed above (Fig. 2). Therefore, the SdH effect in convenient magnetotransport and dHvA effect in magnetic torque and TDO measurements are complementary and provide cross-check for each other.

Earlier studies suggest changes in the topological state with complete substitution of S by Se in $Ni_3In_2S_2$ [8]. To elaborate more on the evolution of the Fermi surface, we performed quantum oscillation studies on two intermediate compounds, $Ni_3In_2S_{1.2}Se_{0.8}$ and $Ni_3In_2S_{0.8}Se_{1.2}$. Here, we utilize torque measurement to study the dHvA effect, which provides consistent results with TDO and resistivity measurements while exhibiting better data quality as shown above in the studies on end compounds $Ni_3In_2S_2$ and $Ni_3In_2Se_2$. The oscillatory component measured at 1.5 K and various magnetic field orientations are shown in Figs. 4(a) and 4(c) for $Ni_3In_2S_{1.2}Se_{0.8}$ and $Ni_3In_2S_{0.8}Se_{1.2}$ respectively. FFT analysis reveals multiple frequencies similar to the end compounds. As shown in Figs. 4(b), with a perpendicular field configuration ($\theta = 0º$, see inset), $Ni_3In_2S_{1.2}Se_{0.8}$ displays three oscillation frequencies of 86 T, 660 T and 1190 T denoted as $F_1$, $F_2$ and $F_3$ respectively. Similarly, three oscillation frequencies have been observed for $Ni_3In_2S_{0.8}Se_{1.2}$ at 114 T ($F_1$), 574 T ($F_2$) and 1292 T ($F_3$) [Figs. 4(d)]. With a rotating magnetic field, the oscillation patterns and frequencies evolve systematically.

To better understand the evolution of Fermi surface from sulfide to selenide in $Ni_3In_2S_{2-x}Se_x$ compounds, we have summarized the angle dependence of the oscillation frequencies for $Ni_3In_2S_2$, $Ni_3In_2S_{1.2}Se_{0.8}$, $Ni_3In_2S_{0.8}Se_{1.2}$ and $Ni_3In_2Se_2$ in Fig. 5(a), (b), (c) and (d) respectively. As shown in Fig. 5(a), oscillation frequencies for $Ni_3In_2S_2$ do not change very significantly with field orientations, indicating a 3D-like Fermi surface. Previous studies on $Ni_3In_2S_2$ suggest that the oscillation frequencies from 0 T – 900 T correspond to the bands with Dirac band crossing along the *T-H-L* in the Brillouin zone, while the higher frequencies originate from the Dirac band crossing along *L-Γ-S* [23]. The Fermi surface corresponding to these bands is shown to be 3D like and agrees with the observed angular dependence of oscillation frequencies in Fig. 5(a). For $Ni_3In_2S_{1.2}Se_{0.8}$ [Fig. 5(b)] and $Ni_3In_2S_{0.8}Se_{1.2}$ [Fig. 5(c)], the frequencies $F_1$ and $F_2$ remain essentially angular-independent and indicate a 3D Fermi surface similar to $Ni_3In_2S_2$. However, $F_3$ in both samples varies strongly with field angle. Assuming a phenomenological model with both 2D and 3D components, the angular dependence for $F_3$ can be well-fitted to $F = F_{3D} + \frac{F_{2D}}{cos\theta}$ [44,40]. This phenomenological model elaborates the angular dependence of the cross-sectional area of the Fermi surface perpendicular to the applied magnetic field. For a 3D spherical Fermi surface, the cross-sectional area perpendicular to the magnetic field does not vary with the field angle $\theta$, resulting in the angle-independent term $F_{3D}$. The angle-dependent term $F_{2D}/cos\theta$ represents a 2D Fermi surface that the cross-sectional area perpendicular to the field orientation varies with $1/cos\theta$. Therefore, when the angular-dependent oscillation frequency fits with the above phenomenological model, such as the case of a corrugated cylindrical Fermi surface, the ratio of $F_{3D}$ and $F_{2D}$ describes the relative weight of 3D and 2D components for the electronic band. The evolution of this weight with composition implies the evolution of dimensionality. The relative weight of 2D and 3D components i.e., $\frac{F_{2D}}{F_{3D}}$, in $Ni_3In_2S_{1.2}Se_{0.8}$ and $Ni_3In_2S_{0.8}Se_{1.2}$ is found to be 1.6 and 2.62, indicating enhanced 2D component with increasing Se content. For $Ni_3In_2Se_2$ the $F_3$ displays a 2D behaviour following $\frac{F_{2D}}{cos\theta}$. Those observations clearly indicate the evolution toward 2D up on Se substitution. Similar behaviour has also been observed in ZrSi(S/Se/Te), in which the electronic structure evolves toward 2D from ZrSiS to ZrSiSe and to ZrSiTe [40].

In Ni$_3$In$_2$S$_2$, the Dirac cone have been probed along $\bar{M}$-$\bar{K}$ direction in the Brillouin zone [7], Unfortunately, the theoretically calculated band structure in earlier reports does not include this Brillouin cut [23,24]. In the recent study [23], the observed high magnetoresistance and high mobility of Ni$_3$In$_2$S$_2$ have been assigned to the Dirac crossing observed along the path *T-H-L*, which corresponds to the lower frequency component of the quantum oscillations. Here, our observation of Fermi surface evolution is corresponding to the higher frequency $F_3$, which motivate us to perform further analysis on this particular band.

In Figs. 6(a)-6(d) we present the magnetic torque oscillation at various temperatures for Ni$_3$In$_2$S$_{2-x}$Se$_x$ (*x* = 0, 0.4, 0.6, and 2) respectively. As we are focussing on the higher frequency band whose dimensionality exhibits clear Se-doping dependence, the high-frequency oscillation component for each sample has been extracted using an FFT filter. As an example, the extracted data at 2 K is shown in the insets of Figs. 6(a)-6(d). The temperature dependence of oscillation amplitude is then obtained for each sample, as shown in Figs. 6(e)-6(h). The oscillation in magnetic torque can be described by the Lifshitz-Kosevich (LK) formula [29,43] with the consideration of the Berry phase term for topological systems [45]:

$$\Delta\tau \propto -B^{1/2} R_T R_D R_S \sin\left[2\pi\left(\frac{F}{B} + \gamma - \delta\right)\right] \quad (1)$$

Where $R_T = \frac{\alpha T \mu}{B\sinh\left(\frac{\alpha T \mu}{B}\right)}$, $R_D = \exp\left(\frac{-\alpha T_D \mu}{B}\right)$, and $R_S = \cos\left(\frac{\pi g \mu}{2}\right)$. Here, $\mu = m^*/m_0$ is the ratio of effective mass *(m$^*$)* and free electron mass *(m$_0$)*. $T_D$ represents the Dingle temperature and $\alpha = \frac{2\pi^2 k_B m_0}{\hbar e}$. The *dHvA* oscillations are described by a sine term with a phase factor $\gamma$–$\delta$, in which $\gamma = \frac{1}{2} - \frac{\phi_B}{2\pi}$ and $\phi_B$ is Berry phase. The factor $\delta$ depends on the dimensionality of the Fermi surface, acquiring a value of $\pm\frac{1}{8}$ for 3D Fermi surface and zero for a 2D one. Therefore, the effective mass *m$^*$* can be obtained by fitting the oscillation amplitude to the temperature-damping term $R_T$ stated above. As shown in Figs. 6(e)-6(h), the fits yield effective masses of

0.74 $m_0$, 0.68 $m_0$, 0.62 $m_0$, and 0.61 $m_0$ for $Ni_3In_2S_2$, $Ni_3In_2S_{1.2}Se_{0.8}$, $Ni_3In_2S_{0.8}Se_{1.2}$, and $Ni_3In_2Se_2$, respectively, which is slightly higher than that obtained from low-field quantum oscillation studies [8,23].

Quantum mobility and the Berry phase are key parameters in quantum oscillation studies on topological materials [29]. Owing to the presence of multiple frequencies, here we obtain these parameters by fitting the oscillation pattern with the LK formula, with the obtained effective mass as a known parameter in the fitting. The oscillation pattern at 2 K for each sample can be well-reproduced by the LK-fitting, as shown in the inset of Fig. 6(a)-6(d). The obtained value of Dingle temperature $T_D$ ranges from 6.91 K to 7.48 K, as summarized in Table 2. Quantum relaxation time can be obtained via $\tau_q = \frac{\hbar}{2\pi k_B T_D}$, ranging from 0.176 ps to 0.162 ps, from which the quantum mobility $\mu_q = \frac{e\tau}{m^*}$ is determined to be within 418.1 cm$^2$/Vs to 466.9 cm$^2$/Vs for various $Ni_3In_2S_{2-x}Se_x$ compounds, as shown in Table 2. Such values are consistent with other topological materials with Kagome lattice such as $Co_3Sn_2S_2$ [46], $ScV_6Sn_6$ [47], $Fe_3Sn_2$ [48], and $TiV_6Sn_6$ [49].

As summarized in Table 2, up on Se substitution, the effective mass and quantum relaxation time reduces systematically, while the quantum mobility increases. Nevertheless, though the dimensionality of the Fermi surface evolves from 3D to 2D, the changes in the values are not significant, implying that the enhanced SOC due to Se substitution may not have a substantial impact on the Fermi surface corresponding to $F_3$ except the dimensionality. From the LK-fit, the Berry phase $\phi_B$ is also obtained, as shown in Table 2. Because of the dimensionality factor $\delta$ in the LK formula, we provide the estimated $\phi_B$ for both the 3D and 2D limit for the intermediate compositions $Ni_3In_2S_{1.2}Se_{0.8}$ and $Ni_3In_2S_{0.8}Se_{1.2}$. The non-trivial Berry phase is consistent with the recent study that predicts topological phase with π-Berry phase in $Ni_3In_2Se_2$ with the presence of SOC [26]. Nevertheless, the earlier experimental study on $Ni_3In_2Se_2$ reveals a trivial Berry phase extracted by using the Landau fan diagram, which has been ascribed to the enhanced SOC that suppresses the topological phase [8]. To fully elucidate the role of SOC in the topological properties of the $Ni_3In_2(S,Se)_2$ system, more experimental efforts are necessary, such as Angle-resolved photoemission spectroscopy along with theoretical calculations which can also clarify the evolution of the dimensionality of the electronic structure. Additionally, in such a substituted system, the formation of possible superstructures, or the rise of chemical short-range order that is characterized by the slight deviation from random distribution of dopants, may also play a role in modifying electronic structures. These factors should be considered together with the role of SOC in determining topological properties.

**Conclusion:**

In summary, we have performed extensive high field quantum oscillations study on $Ni_3In_2S_{2-x}Se_x$ system using torque, TDO and magneto-transport measurements. Our analysis shows that Se substitution in $Ni_3In_2S_2$ results in a gradual evolution of anisotropic Fermi surface. The observed parameters from quantum oscillations analysis suggest that SOC may not strongly modify the electronic states of the higher frequency band. A non-trivial Berry phase is also observed for compositions. These findings open the door for more experimental and theoretical studies on the $Ni_3In_2S_{2-x}Se_x$ system, regarding the evolution of Fermi surface and topological bands.

**Acknowledgment:**

This work was primarily supported by μ-ATOMS, an Energy Frontier Research Center funded by DOE, Office of Science, Basic Energy Sciences, under Award DE-SC0023412. J. S.

acknowledges the support from NIH under award P20GM103429 for XRD. We acknowledge the MonArk NSF Quantum Foundry, which is supported by the National Science Foundation Q-AMASE-i program under the award No. DMR-1906383 for part of resistivity sample fabrication. High-field magneto-transport, TDO and torque measurements were performed at the National High Magnetic Field Laboratory, which is supported by National Science Foundation Cooperative Agreement No. DMR-2128556 and the State of Florida.**Data Availability Statement:**

The data will be made available from the corresponding author upon reasonable request.

**References:**

[1] T.-H. Han, J. S. Helton, S. Chu, D. G. Nocera, J. A. Rodriguez-Rivera, C. Broholm, and Y. S. Lee, Fractionalized excitations in the spin-liquid state of a kagome-lattice antiferromagnet, Nature **492**, 406 (2012).

[2] J.-X. Yin, B. Lian, and M. Z. Hasan, Topological kagome magnets and superconductors, Nature **612**, 647 (2022).

[3] Y. Wang, H. Wu, G. T. McCandless, J. Y. Chan, and M. N. Ali, Quantum states and intertwining phases in kagome materials, Nat Rev Phys **5**, 635 (2023).

[4] T. Ahsan, C.-H. Hsu, Md. S. Hossain, and M. Z. Hasan, Prediction of strong topological insulator phase in kagome metal $RV_6Ge_6$, Phys. Rev. Mater. **7**, 104204 (2023).

[5] S. Okamoto, N. Mohanta, E. Dagotto, and D. N. Sheng, Topological flat bands in a kagome lattice multiorbital system, Commun Phys **5**, 1 (2022).

[6] D. F. Liu et al., Magnetic Weyl semimetal phase in a Kagomé crystal, Science **365**, 1282 (2019).

[7] T. Zhang, T. Yilmaz, E. Vescovo, H. X. Li, R. G. Moore, H. N. Lee, H. Miao, S. Murakami, and M. A. McGuire, Endless Dirac nodal lines in kagome-metal $Ni_3In_2S_2$, Npj Comput Mater **8**, 1 (2022).

[8] L. Cao, G. Liu, Y. Zhang, Z. Yu, Y.-Y. Lv, S.-H. Yao, J. Zhou, Y. B. Chen, and Y.-F. Chen, Crystal growth, transport, and magnetic properties of antiferromagnetic semimetal $Ni_3In_2Se_2$ crystals, Phys. Rev. Mater. **7**, 084203 (2023).

[9] Q. Wang, Y. Xu, R. Lou, Z. Liu, M. Li, Y. Huang, D. Shen, H. Weng, S. Wang, and H. Lei, Large intrinsic anomalous Hall effect in half-metallic ferromagnet $Co_3Sn_2S_2$ with magnetic Weyl fermions, Nat Commun **9**, 3681 (2018).13

**Table 1.** EDS composition and structure parameters obtained from Rietveld refinement of the powder XRD.

| Nominal Composition | EDS Composition | $a$ (Å) | $b$ (Å) | $c$ (Å) | $\alpha$ | $\beta$ | $\gamma$ |
|---|---|---|---|---|---|---|---|
| $Ni_3In_2S_2$ | $Ni_{3.03}In_{1.99}S_{1.97}$ | 5.365(1) | 5.365(1) | 13.541(1) | 90° | 90° | 120° |
| $Ni_3In_2S_{1.2}Se_{0.8}$ | $Ni_{3.05}In_2S_{1.32}Se_{0.62}$ | 5.382(3) | 5.382(3) | 13.748(2) | 90° | 90° | 120° |
| $Ni_3In_2S_{0.8}Se_{1.2}$ | $Ni_{3.1}In_{1.96}S_{0.72}Se_{1.18}$ | 5.394(1) | 5.394(1) | 13.954(1) | 90° | 90° | 120° |
| $Ni_3In_2Se_2$ | $Ni_{3.02}In_{2.01}Se_{1.96}$ | 5.420(6) | 5.420(6) | 14.214(4) | 90° | 90° | 120° |

**Table 2.** Quantum oscillation parameters of $Ni_3In_2S_{2-x}Se_x$.

| Compounds | Frequency (T) | $m^*/m_0$ | $T_D$(K) | $\tau_q$ (ps) | $\mu_q$ (cm$^2$/Vs) | $\phi_B$ |
|---|---|---|---|---|---|---|
| $Ni_3In_2S_2$ | 990 | 0.74 | 6.91(7) | 0.176 | 418.1 | $-0.97\pi$ (3D) |
| $Ni_3In_2S_{1.2}Se_{0.8}$ | 1190 | 0.68 | 6.98(4) | 0.174 | 449.8 | $-1.34\pi$ (3D) <br> $-0.58\pi$ (2D) |
| $Ni_3In_2S_{0.8}Se_{1.2}$ | 1285 | 0.62 | 7.44(2) | 0.163 | 462.2 | $-1.52\pi$ (3D) <br> $-0.78\pi$ (2D) |
| $Ni_3In_2Se_2$ | 1340 | 0.61 | 7.48(1) | 0.162 | 466.9 | $-0.58\pi$ (2D) |

**Figures**

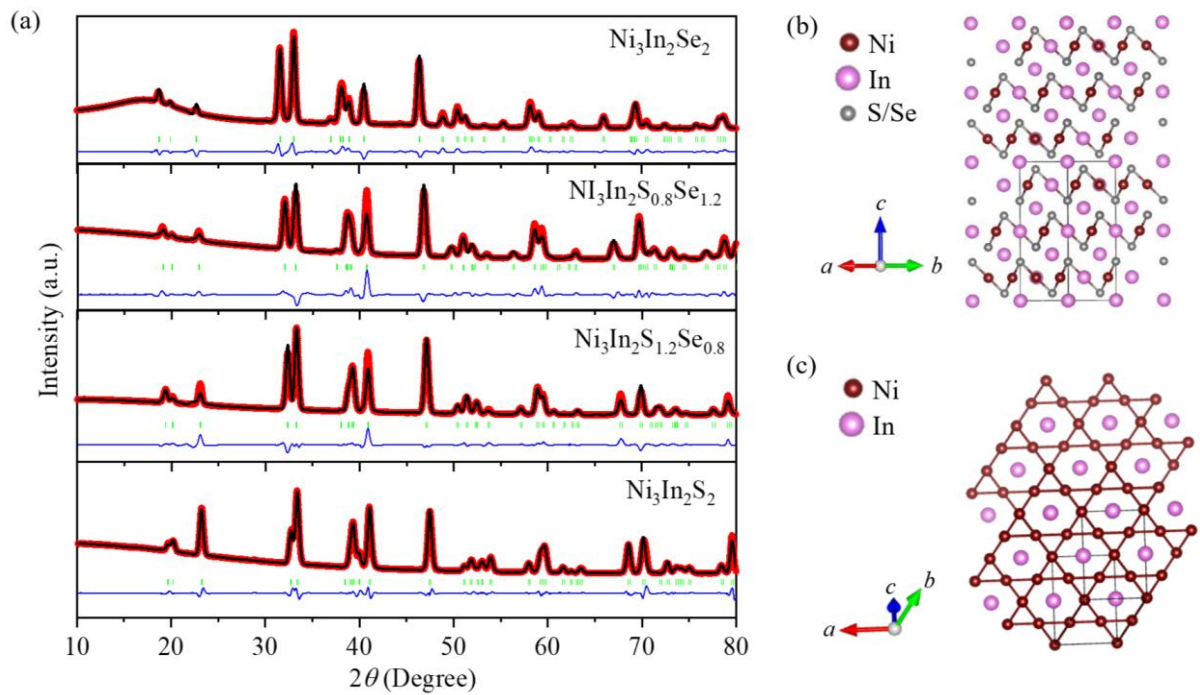

**Fig. 1** (a) Rietveld refined powder XRD pattern of $Ni_3In_2S_{2-x}Se_x$ ($x$ = 0, 0.8, 1.2, and 2) single crystals, in which red symbols denotes the experimental data and the fitted plot is shown by black solid line. Green symbols show Bragg's position and blue solid line shows the difference between experimental and fitted data. (b) Conventional 3D unit cell of $Ni_3In_2(S/Se)_2$. (c) 2D Kagome lattice formed by Ni-In layer in $Ni_3In_2S_{2-x}Se_x$.

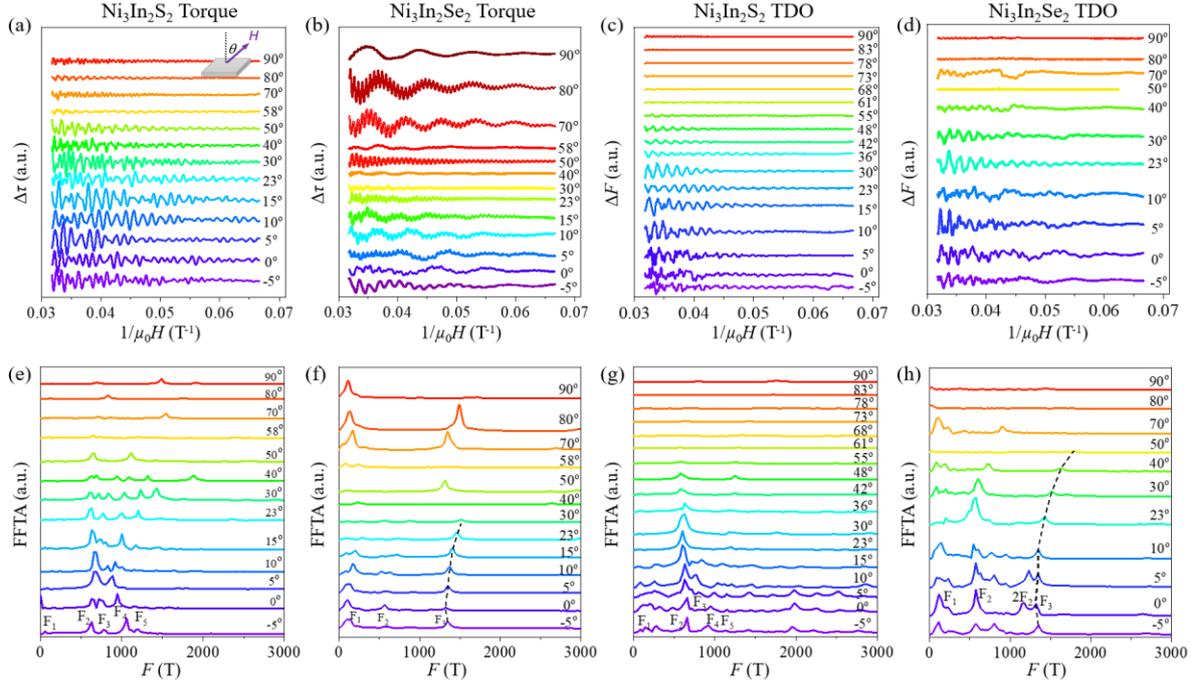

**Fig. 2** (a-b) Oscillatory component of magnetic torque $\Delta\tau$ measured at 1.5 K at different magnetic field orientations for (a) $Ni_3In_2S_2$ and (b) $Ni_3In_2Se_2$. Inset in (a): magnetic field orientation with respect to the crystal plane. (c-d) Oscillatory component of TDO measured at 1.5 K at different magnetic field orientations for (c) $Ni_3In_2S_2$ and (d) $Ni_3In_2Se_2$. (e-f) FFT spectra of the torque oscillation for (c) $Ni_3In_2S_2$ (d) $Ni_3In_2Se_2$. (g-h) FFT spectra of the TDO oscillation for (g) $Ni_3In_2S_2$ (h) $Ni_3In_2Se_2$.

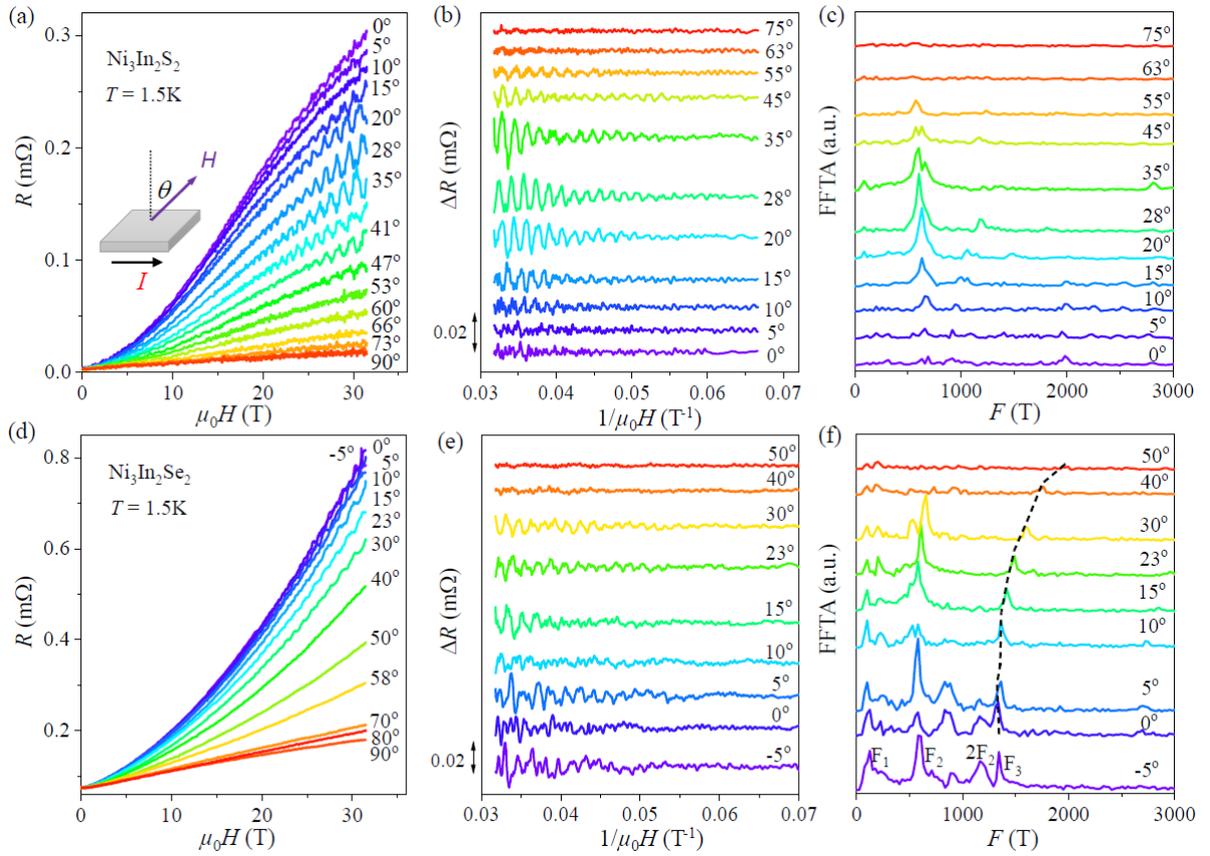

**Fig. 3** Magnetic field dependence for resistance (*R*) at 1.5 K measured under different magnetic field orientations for (a) $Ni_3In_2S_2$ and (d) $Ni_3In_2Se_2$. The extracted oscillatory components are shown in (b) $Ni_3In_2S_2$ and (e) $Ni_3In_2Se_2$. FFT spectra for the oscillatory component ΔR for (c) $Ni_3In_2S_2$ and (f) $Ni_3In_2Se_2$.

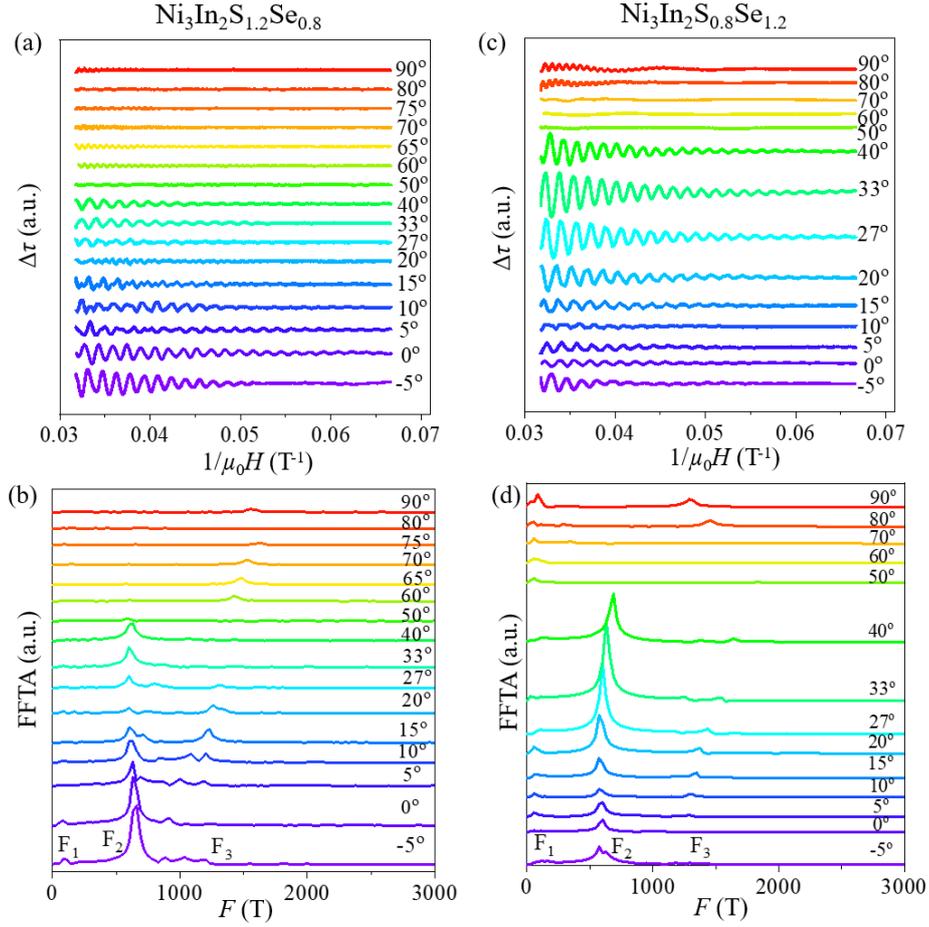

**Fig. 4** dHvA oscillations in torque measurements at 1.5 K measured under various magnetic field orientations for (a) $Ni_3In_2S_{1.2}Se_{0.8}$ and (c) $Ni_3In_2S_{0.8}Se_{1.2}$. FFT spectra for the oscillatory component $\Delta\tau$ for (c) $Ni_3In_2S_{1.2}Se_{0.8}$ and (d) $Ni_3In_2S_{0.8}Se_{1.2}$.

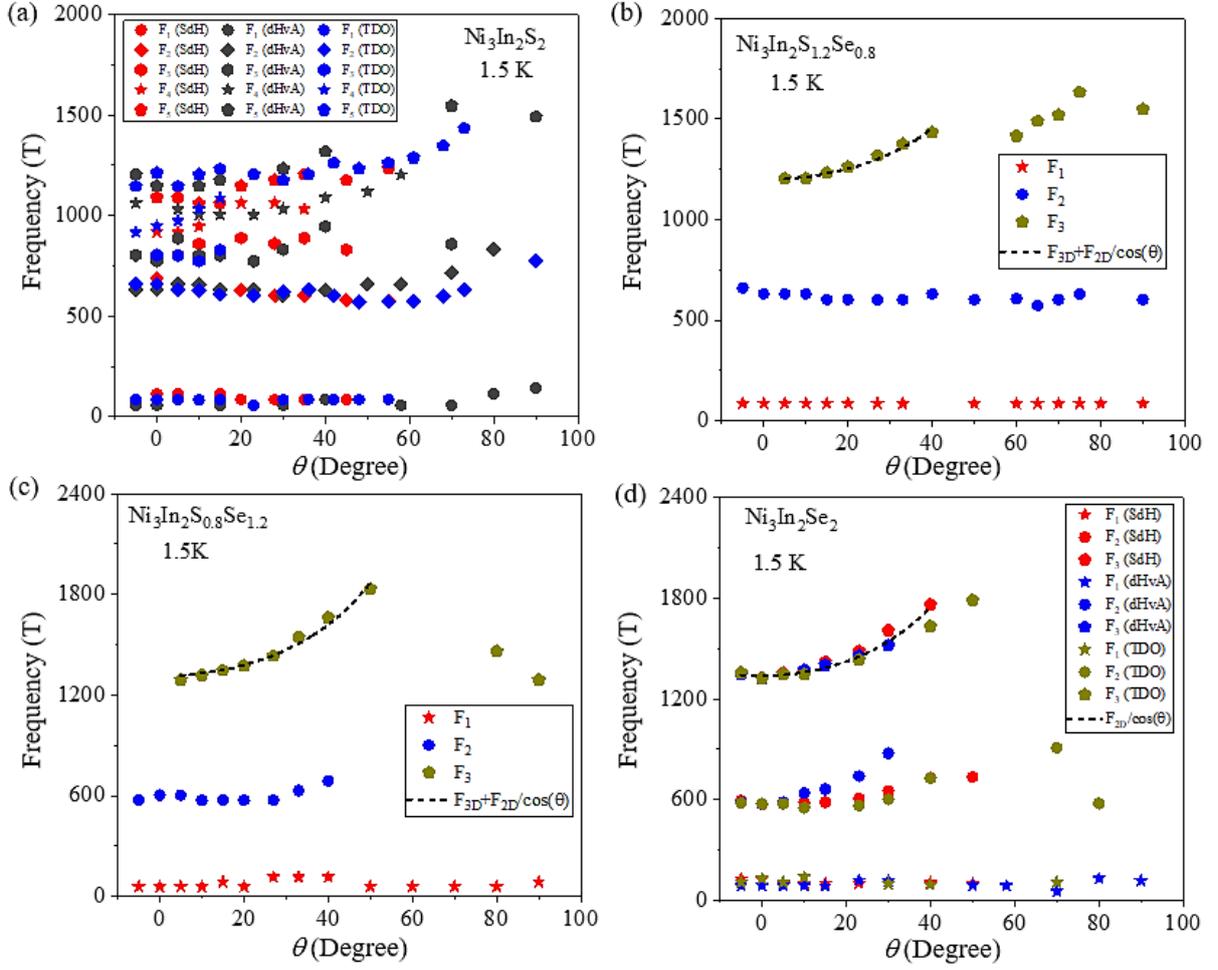

**Fig. 5** Angular dependence of the oscillation frequencies for (a) $Ni_3In_2S_2$, (b) $Ni_3In_2S_{1.2}Se_{0.8}$, (c) $Ni_3In_2S_{0.8}Se_{1.2}$, and (d) $Ni_3In_2Se_2$. The black dashed lines in (b) and (c) show the fitting for $F_3$ frequency $F_3$ with $F = F_{3D} + \frac{F_{2D}}{\cos\theta}$. The black dashed lines in (d) show the fitting for $F_3$ frequency $F_3$ with $F = \frac{F_{2D}}{\cos\theta}$.

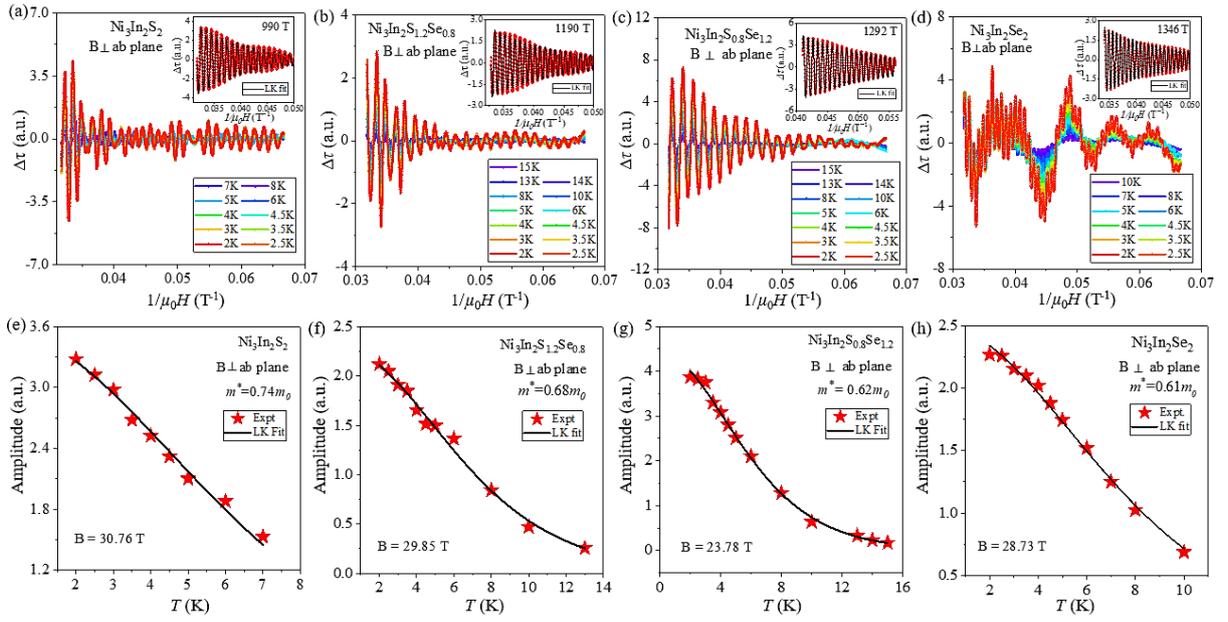

**Fig. 6** (a-d) dHvA oscillations observed in torque measurements performed at different temperatures under out-of-plane magnetic field direction for (a) $Ni_3In_2S_2$, (b) $Ni_3In_2S_{1.2}Se_{0.8}$, (c) $Ni_3In_2S_{0.8}Se_{1.2}$, and (d) $Ni_3In_2Se_2$. The insets show the high-frequency oscillatory components at 2 K, together with LK fitting (black line). (e-h) Temperature dependence of oscillation amplitude for the high-frequency oscillatory components for (e) $Ni_3In_2S_2$, (f) $Ni_3In_2S_{1.2}Se_{0.8}$, (g) $Ni_3In_2S_{0.8}Se_{1.2}$, and (h) $Ni_3In_2Se_2$. The black line in each panel shows the fitting to extract effective mass (see text).